\documentclass[runningheads]{llncs}
\usepackage{graphicx}
\usepackage[noend]{algpseudocode}
\usepackage{bm}
\usepackage{algorithm}
\usepackage{algpseudocode}
\usepackage{amsmath}
\usepackage{graphics}
\usepackage{epsfig}
\usepackage{graphicx}
\usepackage{subfigure}
\usepackage{multirow}
\usepackage[normalem]{ulem}
\useunder{\uline}{\ul}{}
\newcommand{\eat}[1]{}
\usepackage{makecell}
\usepackage{balance}
\usepackage{diagbox}
\usepackage[misc]{ifsym}

\begin{document}
\title{Nearest Neighbor Classifier with Margin Penalty for Active Learning}
\toctitle{Nearest Neighbor Classifier with Margin Penalty for Active Learning}
\titlerunning{Nearest Neighbor Classifier with Margin Penalty for Active Learning}
%
\author{Yuan Cao\inst{1,2,3} \and
Zhiqiao Gao\inst{2} \and
Jie Hu\inst{2} \and
Mingchuan Yang\inst{2} \and
Jinpeng Chen\inst{1,3}\Letter
}
\tocauthor{Yuan Cao, Zhiqiao Gao, Jie Hu, Mingchuan Yang, Jinpeng Chen}
\authorrunning{Y. Cao et al.}
%
\institute{School of Computer Science (National Pilot Software Engineering School), Beijing University of Posts and Telecommunications, Beijing, China\\
\email{\{caoyuanboy, jpchen\}@bupt.edu.cn}\and
China Telecom Corporation Limited Research Institute, Beijing, China
\email{\{gaozhq6,hujie1,yangmch\}@chinatelecom.cn}\\ \and
Key Laboratory of Trustworthy Distributed Computing and Service (BUPT), Ministry of Education, Beijing, China}
\maketitle              
\begin{abstract}
As deep learning becomes the mainstream in the field of natural language processing, the need for suitable active learning method are becoming unprecedented urgent. \eat{the need for large amounts of annotated data has become the biggest pain point for deep learning. As text classification tasks often require datasets with large amount of long text, annotating text data is more labour-intensive than in other domains. Therefore, the need for suitable active learning methods are unprecedented urgent in the field of natural language processing. } Active Learning (AL) methods based on nearest neighbor classifier are proposed and demonstrated superior results. However, existing nearest neighbor classifiers are not suitable for classifying mutual exclusive classes because inter-class discrepancy cannot be assured. As a result, informative samples in the margin area can not be discovered and AL performance are damaged. To this end, we propose a novel \textbf{N}earest neighbor \textbf{C}lassifier with \textbf{M}argin penalty for \textbf{A}ctive \textbf{L}earning(NCMAL). Firstly, mandatory margin penalties are added between classes, therefore both inter-class discrepancy and intra-class compactness are both assured. Secondly, a novel sample selection strategy is proposed to discover informative samples within the margin area. To demonstrate the effectiveness of the methods, we conduct extensive experiments on three real-world datasets with other state-of-the-art methods. The experimental results demonstrate that our method achieves better results with fewer annotated samples than all baseline methods.

\keywords{Active Learning \and Text Classification \and Bert}
\end{abstract}

\section{Introduction}

\begin{figure}
    \centering
    \includegraphics[width=0.5\linewidth]{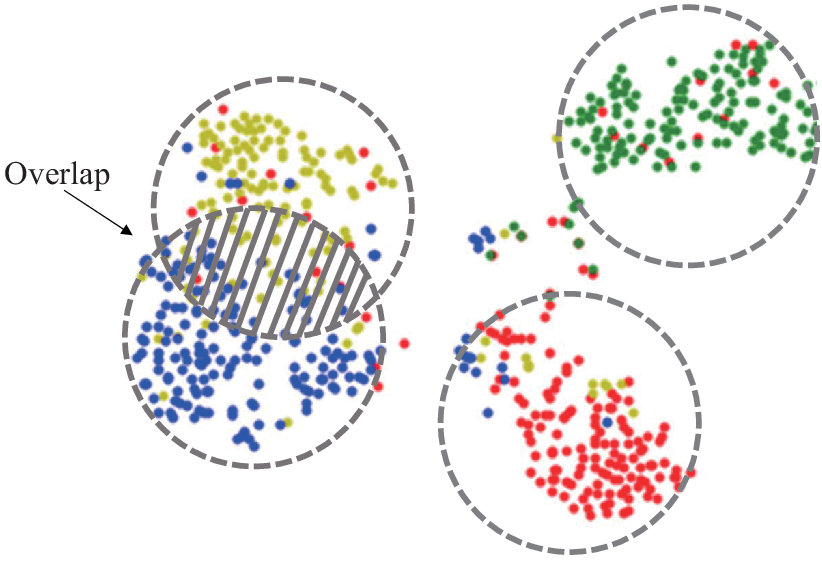}
    \caption{Visualization of sample distribution results after 10 rounds of AL using NCENet on the \textbf{AGNEWS} dataset\eat{ There is a clear overlap between the yellow class and the blue class, which indicates that the inter-class discrepancy cannot be assured by NCENet. As a result, the yellow class and the blue class are mixed together and become not exclusive. No clear classification boundary or margin area can be found in the learned sample distribution space, thus damaging both classification performance and AL performance.}}
    \label{fig:ncebad}
\end{figure}

\eat{Recently, Deep Learning (DL) has shown unparalleled ability in many areas especially in the field of natural language processing (NLP). The presence of BERT\cite{devlin2018bert} has changed the landscape of NLP and BERT has achieved state-of-the-art performance in nearly all sub-fields in NLP. However, compared with traditional machines learning algorithms, DL's superb learn capabilities relies heavily on large amount of labeled data. As a result DL encounters greater problems in the data collection and processing phase.

\eat{Compared with the fiery DL, Active Learning(AL) doesn't receive much interest from researchers before the rise of DL, because machine learning does not require such amount of labeled data.} In recent years, riding on the rise of DL, Active Learning (AL) is gradually receiving more attention\cite{dor2020active}\cite{prabhu2021multi}\cite{nguyen2021information}\cite{li2021unsupervised}\cite{zhou2021mtaal}. The goal of AL is to maximize model performance while minimize labeling costs, which may be a possible solution for DL models that eager for large amount of labeled data and may help ease the data shortage problems of DL. }

Recently, Deep Learning (DL) has shown unparalleled ability in many areas especially in the field of natural language processing (NLP). DL-based\cite{devlin2018bert}\cite{lai2015recurrent}\cite{lan2019albert} text classification methods has changed the landscape of text classification and achieved state-of-the-art performance. However, DL's superb learn capabilities havily relies on large amount of labeled data. As a result, active learning (AL), which aims to maximize model performance while minimize labeling costs, is gradually receiving more attention\cite{dor2020active}\cite{prabhu2021multi}\cite{nguyen2021information}\cite{li2021unsupervised}\cite{zhou2021mtaal}, and may help ease the data shortage problems of DL.

So far, many AL methods analyse the output logits of the traditional softmax classifier for sample selection. The uncertainty-based method\cite{RenXCHLGCW22}\cite{gal2017deep}\cite{dor2020active}, a bunch of AL methods whose presence may date back to the era of machine learning, aims to calculate the uncertainty of the output logits to select the most uncertain samples for the model. Intuitively, those uncertainty-based methods are inherited in deep learning models.\eat{ However, the deep model may be overconfident in the final output logits. Previous work by DBAL\cite{gal2017deep} has made this point and introduces Bayesian inference into DL by using Monte-Carlo Dropout in the testing phase in order to obtain more accurate uncertainty scores. The newly obtained uncertainty score is then used to select samples for labeling.} However, these ported methods didn't perform as well as they do on machine learning.

\eat{In another direction, }Fang et al.\cite{wana2021nearest} pointed out that the problem encountered with deep models is actually a "false generalize" problem. DL models learn softmax classification boundaries form labeled samples, and incorrectly generalizes the classification boundaries to unlabeled samples. Fang then discards the traditional softmax classifier structure and utilizes a soft nearest neighbour classifier that classifies target samples by selecting prototype vectors on the feature space. \eat{Two sample selection strategies are also proposed on the basis of NCENet, which eventually has excellent performance in both image classification and object detection .} NCENet\cite{wana2021nearest} is proposed to avoid the "false generalize" problem by complete abandonment of the softmax classifier structure.

However, NCENet also has its shortcomings. The main structure of NCENet consists of $n$ sigmoid functions instead of one softmax function. Although the sigmoid structure can be used in multi-classification scenario, it may encounter difficulty with classes that are mutually exclusive. For example, Fig.\ref{fig:ncebad} shows a typical scene from a real AL training process on the \textbf{AGNEWS} dataset, visualised by t-SNE\cite{van2008visualizing}. A clear overlap can be easily seen from the yellow class and blue class. The yellow class and the blue class are mixed together and no clear classification boundary can be established. In this way, two classes that were mutually exclusive become non-mutually exclusive. These are important indications that inter-class differences are not guaranteed by the sigmoid classifier. We define this phenomenon as the "non-exclusive problem". Solving the "non-exclusive problem" will enhance the performance of the model in classification and AL scenarios.\eat{An intuitive solution is that, by adding mandatory inter-class margin and creating a margin area between two neighbor classes, the inter-class discrepancy can be assured, so that the "non-exclusive problem" can be solved. When it comes to the AL scenario, we believe that samples in the margin area contains enormous information for the model to learn, which may be the key to enhance AL performance.}

To this end, we propose \textbf{N}earest neighbor \textbf{C}lassifier with \textbf{M}argin penalty for \textbf{A}ctive \textbf{L}earning(NCMAL), which ensures class not overlapping by adding mandatory margins between classes so that the sigmoid classifier can be used in classifying mutual exclusive classes. In other words, inter-class discrepancy can be assured with the mandatory inter-class margin added. And at the same time, as we project the whole feature space onto a n-dimensional hyper-sphere, higher inter-class discrepancy brings higher intra-class compactness. As a result, the classification accuracy can be improved. Meanwhile, with margin area added, we believe that unlabeled samples within the margin area has a relatively high uncertainty, and have a high priority when labeling. We proposed a sample selection strategy that focuses high priority samples in the margin area.

Our contributions are summarized as follows:
\vspace{-\topsep}
\begin{itemize}
  \item The proposed NCMAL effectively increases the inter-class discrepancy and make sigmoid-based classifier suitable for classifying mutual exclusive classes.
  \item We prove samples in the margin area tend to be more informative and propose a special sample selection strategy, which gives high priority to samples in the margin area.
  \item NCMAL outperforms state-of-the-art AL methods for text-classification on three real-world datasets.
\end{itemize}
\vspace{-\topsep}

\section{Related Work}

We focus on AL in pool-based scenarios.\eat{In pool-based AL scenario, the learner has the access to all the samples from the unlabeled set. And during each AL iteration, $k$ samples are selected from the unlabeled set for labelling and added to the labeled set, where $k$ is often called budget.} Pool-based AL methods can be roughly divided into three categories: uncertainty-based, representation-based and fusion methods which combines uncertainty-based method and representation-based method.

\textbf{Uncertainty-based Method.} AL has been of interest to researchers since the days of machine learning, when one wanted to obtain better model performance with fewer labelled samples. In the most intuitive way of thinking, one determines whether a sample needs to be labelled by the uncertainty of the model on the sample. Different methods\cite{nguyen2022bayesian}\cite{nafa2022active} have different measures of uncertainty, such as \cite{culotta2005reducing}\cite{settles2008analysis}\cite{settles2009active} based on least model confidence, \cite{scheffer2001active} based on margin sampling, and \cite{lewis1994sequential}\cite{zhu2008active} by measuring the entropy of the probability distribution to determine the uncertainty of this classification. In addition, \cite{gal2017deep} introduces Bayesian inference through the use of a Monte Carlo Dropout, which measures sample uncertainty more accurately by enabling the Dropout in the testing phase. However, the computational efficiency of this method is greatly limited by the need to perform multiple forward propagation.

\textbf{Representation-based Method.} Representation-based methods aims to select the most important samples for labelling by analyzing the distribution of the unlabelled samples. As in the DAL (Discriminative Active Learning)\cite{gissin2019discriminative} method, a binary classifier is trained to discriminate whether a sample comes from the labelled set or the unlabelled set, so that samples that best represent the entire data set can be selected. The EGL(Expected Gradient Length)\cite{huang2016active} method measures the impact of a sample on the model by calculating the EGL of the labeled sample, and selects the labeled sample based on this criterion. Coreset\cite{sener2017active} is also an emerging and very effective method that models the entire AL process as a coreset problem. By solving the corresponding coreset problem in the learned representation space, samples that can best represent the entire dataset are selected to be labeled.

\textbf{Fusion Method.} In addition, there are many methods that combine the uncertainty-based method with the representation-based method, e.g. the BADGE\cite{ash2019deep} method can be considered as a combination of the BALD\cite{houlsby2011bayesian} method and the Coreset\cite{sener2017active} method, and has been experimented on several models. For example, LL4AL\cite{yoo2019learning} uses an additional network structure to predict the "loss" of a sample, which gives a more accurate measure of the diversity and uncertainty of the sample, and the top $L$ labeled samples are obtained by sorting the loss values in descending order. \eat{In addition to this, there a a bunch of adversarial AL methods, which utilizes different adversarial network structures to achieve AL goal. Typical methods like Wasserstein Adversarial AL(WAAL)\cite{shui2020deep}, which utilizes Wasserstein distance as measurements of distances. Variational Adversarial Active Learning(VAAL)\cite{sinha2019variational} and its expanded method Task-Aware Adversarial Active Learning(TA-VAAL)\cite{kim2021task} both uses Variational Autoencoder(VAE)\cite{kingma2013auto} to learn the latent representation space of the data, while TA-VAAL has a additional loss prediction module. }The NCENet method uses a Nearest Neighbor Classifier to replace the traditional softmax classifier, thus solving the "false generalize" problem of the softmax classifier.

\eat{
There are many existing methods that can increase inter-class discrepancy. Most of these methods originate from the field of face recognition. Centre loss\cite{wen2016discriminative} adds penalties between features and corresponding class centres and then increases intra-class compactness. SphereFace\cite{liu2017sphereface} first introduced the idea of angular margin and its successor CosFace\cite{wang2018cosface} adds cosine margin penalty to the logits. Arcface\cite{deng2019arcface}, compared with methods described above, has a constant linear angular margin throughout the whole interval. Those methods are all effective when dope with inter-class discrepancy problem.}

\section{Methodology}
In this section, we first introduce the NCMAL in detail. Then, the sample selection strategy, which aim to informative samples from the margin area, is described.

\begin{figure}[htbp]
    \centering
    \includegraphics[width=\linewidth]{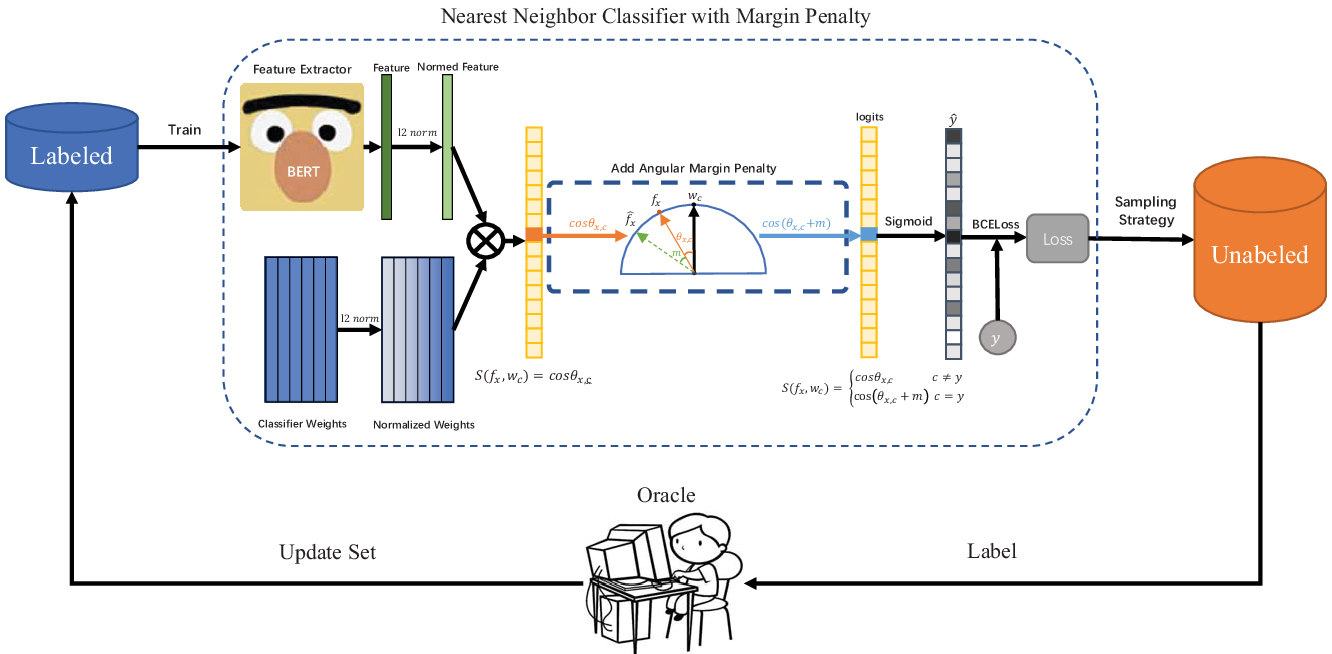}
    \caption{One AL iteration of NCMAL}
    \label{fig:addmargin}
\end{figure}

\subsection{Overall Framework}

Fig.\ref{fig:addmargin} shows the whole structure of NCMAL. This work is applied to pool-based active learning scenarios. Specifically, the algorithm is initialized with a small set of labeled samples $\mathcal{L}$ and a larger set of unlabeled samples $\mathcal{U}$. The samples $x_i \in \mathcal{L}$ all have corresponding labels $y_i$, while the unlabeled samples $x_i \in \mathcal{U}$ have no labels. Using $\mathcal{L}$ as training data, a text classifier $g(x|\Theta):\mathbf{X}\rightarrow\mathbf{Y}$ is trained. The goal of the sample selection strategy is to select $K$ samples from $\mathcal{U}$ by the classification result of the trained model $g(x|\Theta)$. The selected $K$ samples are then annotated and added to $\mathcal{L}$, and used as the training data of the next round of training. The whole algorithm can be summarized as Algorithm \ref{alg:Framework}.

\eat{\subsection{Nearest Neighbor Classifier}

In NCENet, the trained classification model consists mainly of a feature extraction network and a Nearest Neighbor Classifier. Firstly it selects $N$ prototype vectors $m_{c, n}$ ($n=1,...,N$ for each class $c \in C$) in the feature space by learning from the labelled samples. For simplicity we set $N=1$. The distance between the sample feature $f_x$ extracted by the feature extraction network and the prototype vector corresponding to the category represents the similarity between the sample and the class. The similarity between the sample feature and the prototype vector corresponding to the class is defined as
\begin{equation}
 S(f_x,  w_c) = -d(f_x, m_c)   
\end{equation}
where $m_c$ is the prototype vector corresponding to the category $c$. The values of $m_c$ are randomly initialised and gradually optimised during the training process by a gradient descent method. $d(\cdot)$ is a distance function, defined in the original text as either the Euclidean distance or the cosine distance.
Accordingly, a binary cross entropy loss can be used as the Loss function of the NCENet, i.e.:
\begin{equation}
\label{L1}
L_{1}=-\sum_{c}{y_{x,c}\log p_c(f_x)} + (1-y_{x,c})\log(1-p_c(f_x))
\end{equation}
where $y_{x,c} = 1$ if the sample $x$ belongs to the category $c$ and vice versa $y_{x,c} = 0$.}

\begin{algorithm}[htb] 
\caption{\emph{Nearest Neighbor Classifier with Margin Penalty for Active Learning(NCMAL)}} 
\label{alg:Framework} 
\begin{algorithmic}[1] 
\Require 
Unlabeld set $\mathcal{U}$, initial budget $K_{init}$, budget $K$, margin factor $m$, deflation factor $s$, AL rounds $r$.
\Ensure 
Model parameters $\Theta$, labeled set $\mathcal{L}$
\State Initialize $\Theta$ from Normal Distribution $\mathcal{N}(0, 0.01)$;
\State $\mathcal{L} \longleftarrow Random\_Select\_K\_Sample\_From(\mathcal{U}, K_{init})$
\For {$i = 1, 2, ..., r$} 
    \For {$x_i, y_i \in \mathcal{L}$}
    \State Compute $o_{x_i,c}$ for every $c \in C$ according to Eq. \ref{eq:output};
    \State Compute loss $L$ according to Eq. \ref{eq:loss}
    \State Update parameter $\Theta$ by gradient decent optimization with loss $L$
    \EndFor
    \State \textbf{end for}
    \For {$x \in \mathcal{U}$}
    \State Compute Margin Confidence score $C^{Margin}_x$ according to Eq. \ref{eq:margin}
    \EndFor
    \State \textbf{end for}
    \State $\mathcal{L}_i \longleftarrow Top\_K\_Sample\_Selection\_By\_Confidence\_Score(Conf(\mathcal{U}), K)$
    \State $\mathcal{L} \longleftarrow \mathcal{L}+\mathcal{L}_i$; \State $\mathcal{U} \longleftarrow \mathcal{U}-\mathcal{L}_i$
\EndFor 
\State \textbf{end for}
\State \Return $\Theta$,$\mathcal{L}$;
\end{algorithmic} 
\end{algorithm}

\subsection{Nearest Neighbor Classifier with Margin Penalty}

\eat{NCE converts the original $n$-class multi-classification task into $n$ binary classification tasks, where the probability of a sample belonging to a class is directly related to the distance between the sample feature vector and the prototype vector. That is, the performance of the whole NCENet depends heavily on the position of the prototype vector in the feature space. In NCENet, however, the prototype vectors are trained by relying only on gradient descent. As mentioned earlier, the prototype vectors tend to have a small spacing in practice, especially if the corresponding two classes are similar in terms of features. In the actual training it will lead to a situation where the boundaries of the two classes overlap, thus leading to a blurred inter-class spacing, and these otherwise highly informative points cannot be well selected in the subsequent sample selection.
In order to solve this problem in NCENet, naturally, we refer to the approach in \cite{deng2019arcface} and add an angular margin penalty between classes during training, which can increase the inter-class sample spacing and reduce the intra-class compactness.}

In existing nearest neighbor classifier methods\cite{wana2021nearest}\cite{DBLP:journals/jmlr/KontorovichSU17}, take NCENet as an example, the classification result of an arbitrary sample mainly depends on the similarity between the feature vector $\bm{f}_x$ and the prototype vector $\bm{w}_c, c\in C$. The feature vector is extracted by a arbitrary feature extraction network (e.g. Bert, TextCNN). The output score function can be written as, 
\begin{equation}
\label{eq:output1}
    o_{x, c} = \sigma(\overline{S}(\bm{f}_x, \bm{w}_c))
\end{equation}
where $\overline{S}(\cdot,\cdot)$ is an arbitrary similarity function. As we can see from Eq.\ref{eq:output1}, the main structure of the NCENet consists of $n$ sigmoid classifiers.
\eat{By utilizing $n$ sigmoid classifiers, NCENet transforms the $n$-class multi-classification problem into $n$ binary classification problem. \eat{However, this kind of transformation suits for classes that are not mutual exclusive. And as the prototype vectors are trained by relying only on gradient descent they tend to have a small spacing in practice, especially when the corresponding two classes are similar in features.}However, as we mentioned before, along with the conversion comes the "non-exclusive" problem.}

As we mentioned before, the transformation from softmax to n-sigmoid brings the "non-exclusive" problem. In order to solve this problem, intuitively, we refer to the approach in \cite{deng2019arcface} and add an angular margin penalty between classes during training, which can increase the inter-class discrepancy and the intra-class compactness. In AL scenario, samples in the overlapping area contains much more information to better separate the overlapped classes. By adding mandatory margin, overlapping areas are now margin areas. Samples used to be in the overlapping area are now located in the created margin area, and can be better measured by the special sample selection strategy we describe later.

First, we utilize dot product to measure the similarity between vectors. We define the similarity between a feature vector $\bm{f}_x$ and prototype vector corresponding to class $c$ as
\begin{equation}
    S(\bm{f}_x, \bm{w}_c) = \bm{f}_x^\mathrm{T}\bm{w}_c = ||\bm{f}_x|| ||\bm{w}_c||cos\theta_{x,c}
\end{equation}
where $\theta_{x,c}$ is the angle between $\bm{f}_x$ and $\bm{w}_c$. We apply $l_2$ regularization to $\bm{w}_c$ so that $||\bm{w}_c||=1$. We also regularise $\bm{f}_x$ and rescale to $s$. With $l_2$ regularisation, we project $\bm{f}_x$ and $\bm{w}_c$ onto a feature space shaped as a hypersphere with radius $s$, making the multi-classification prediction dependent only on the angle between the sample vector and the prototype vector. The similarity can be then described as
\begin{equation}
    \label{eq:distance}
    S(\bm{f}_x, \bm{w}_c) = s*cos\theta_{x,c}.
\end{equation}
Since the sample features as well as the prototype vectors are projected onto the same hypersphere with radius $s$, adding a angular margin becomes possible. We add a angular margin penalty $m$ to $\theta_{x,y}$, where $y$ is the ground-truth label of sample $x$. The new similarity function of $\bm{f}_x$ and $\bm{w}_y$ can be written as
\begin{equation}
    S(\bm{f}_x, \bm{w}_c) = s*cos(\theta_{x,y} + m)
\end{equation}
The whole similarity function can be written as
\begin{equation}
    \label{eq:similarity}
    S(\bm{f}_x, \bm{w}_c) = \left\{ \begin{array}{lr}
        s*cos\theta_{x,c} & c \neq y \\
        s*cos(\theta_{x,c} + m) & c = y
    \end{array}\right.
\end{equation}

\eat{where $\theta_{x,y}$ is the angle between $\bm{f}_x$ and real class prototype vector $w_y$.} We then apply sigmoid classifier to the similarity score in order to calculate the probability of sample $x$ belong to class $c$. 
\begin{equation}
    \label{eq:output}
    o_{x,c} = \sigma(S(\bm{f}_x, \bm{w}_c))
\end{equation}
A binary cross entropy loss function is applied. The loss function can be rewritten as\
\begin{equation}
\label{eq:loss}
\begin{aligned}
L  &=-\sum_{c}{y\log o_{x,c}+(1-y)\log(1-o_{x,c})}\\
    &= -\log \sigma(s*cos(\theta_{x,y}+m))-\sum_{c\neq y}{\log(1-\sigma(s*cos\theta_{x,c}))}
\end{aligned}
\end{equation}
And in the testing phase, the sample will be predicted to be the class with maximum Eq. \ref{eq:distance}.

\subsection{Sample Selection}
In each round of active learning, we rely on the probability output $o_{x,c}$ to select the samples. As we mentioned in the previous section, the NCMAL creates a margin area between classes, and we believe that samples located in the margin area shall have higher priority when labeling. To best find the samples in the margin area, we proposed a confidence score function for NCMAL in order to give samples closer to the margin area a relatively high confidence score.

\eat{To discover samples within the margin area, the most intuitive way is to find the difference between the largest and second largest predicted probability. The smaller the difference is, the closer the sample is to margin areas.}

\textbf{\textit{Margin Confidence}}
\begin{equation}
\label{eq:margin}
    C^{Margin}_x = -|o_{x,c_0} - o_{x,c_1}|,
\end{equation}
where $c_0, c_1$ are the classes with largest and second largest output probabilities respectively. It is worth noting that the Margin confidence score is closely related to the difference in hyperarc length from the sample point to the two nearest class centers after projected onto the feature hypersphere.

The higher the confidence score is, the higher priority the sample obtains when labeling. Samples with top-$k$ confidence score will be queried and manually labeled for the next AL iteration. The effects of different sampling strategies will be discussed in the experimental subsection. \eat{The effects of different sampling strategies will be discussed in the experiment section.}

\section{Experiments}
NCMAL is tested on several datasets on a text classification task. In this section, we describe the implementation and results of the experiments in detail.
\subsection{Experimental Settings}

\ \ \ \ \textbf{\textit{Datasets.}} Three different text classification datasets are used to prove the effectiveness of our method. The three datasets consists of two public datasets and one private dataset (\textbf{Telecom}). The \textbf{Telecom} dataset comprises of a total of 7,302 real-word messages from Chinese users. These messages were labeled and divided into 25 pre-defined mutually exclusive classes by a dedicated team. The dataset was divided into 5,841 training samples and 1,461 test samples. The Statistics for these three datasets are shown in Table \ref{tab:ds-stat}.
\begin{table}[]
\centering
\caption{Statistics of datasets}
\label{tab:ds-stat}
\begin{tabular}{l|l|l|l|l}
\hline
Dataset               & \textbf{AGNEWS}      & \textbf{IMDb} & \textbf{Telecom}\\ \hline\hline
\#class                 & 4         & 2     & 25\\ \hline
\#train                 & 120000    & 25000  & 5841   \\ \hline
\#test                  & 7600      & 25000 & 1461     \\ \hline
\#init budget           & 50        & 100   & 500     \\ \hline
\#budget               & 10         & 20    & 20     \\ \hline
\#round               & 50         & 50     & 30     \\ \hline
\end{tabular}
\end{table}

\textbf{\textit{Feature Extraction Network.}}
The commonly used pre-trained Bert\cite{devlin2018bert} were chosen as the Feature Extraction Network. For all AL methods, hyperparameters were chosen consistently for fairness considerations. Necessary changes are made on the original Bert structure due to the requirements of both NCENet and NCMAL, while the number of parameters remains the same for fairness considerations.
\setcounter{footnote}{0}

\textbf{\textit{Trainning Details.}}
NCMAL is implemented on Pytorch and trained on 4* NVIDIA Tesla V100. Both init-budget and budget selection on different datasets is shown in Table \ref{tab:ds-stat}. Batch size was set to 10 and the model trained on a learning rate of $2e^{-5}$ using the AdamW\cite{loshchilov2017decoupled} optimizer. For each AL sampling method, 10 different random number seeds are used for testing and the final performance of the method was averaged over 10 experiments.

\textbf{\textit{Baselines.}} We compare our approach to the following baselines.
\vspace{-\topsep}
\begin{itemize}
\item \textbf{Random.} It is the most commonly used baseline in active learning. The samples added to the labeled set in each round are randomly selected from the unlabeled set.
\item \textbf{DBAL(Deep Bayesian Active Learning).} Monte Carlo Dropout was used to provide a more accurate measure of the uncertainty of the classifier. Both Confidence and Entropy were used in the final sample selection stage and the best performing of the two methods was selected as the performance of this method in the end.
\item \textbf{Coreset.} Samples that best cover the entire feature space are selected. We chose two implementations of Coreset as described in \cite{sener2017active}, the greedy version of Coreset are implemented.
\eat{\item \textbf{EGL-Word.} Samples with top-$k$ Expected gradient norm on the Word Embedding layer are selected for labelling. Expected Gradient Length(EGL) is calculated by mathematically expecting the probability of the classifier output with the corresponding back-propagation gradient value.}
 \item \textbf{BADGE\cite{ash2019deep}.} It can be viewed as a combination of EGL and Coreset, and ensures diversity and uncertainty at the same time.
 \item \textbf{NCENet.} We implemented NCENet as described in \cite{wana2021nearest}.
 \end{itemize}
The implementation\footnote{https://github.com/GhostAnderson/Nearest-Neighbor-Classifier-with-Margin-Penalty-for-Active-Learning} was based on the code\footnote{https://github.com/dsgissin/DiscriminativeActiveLearning} made available by \cite{gissin2019discriminative}.
 
\vspace{-\topsep}
\subsection{Model Effect}

\begin{table}[]
\centering
\caption{Accuracy performance on full training scenario. The best performing method in each row is boldfaced}
\label{tab:ful-tr}
\begin{tabular}{l|l|l|l|l}
\hline
\diagbox{Dataset}{Classifier}               & Softmax      & NCENet  &  NCMAL\\ \hline\hline
\textbf{AGNEWS}                 & 94.42\%         & 94.67\%     & \textbf{94.87}\%    \\ \hline
\textbf{IMDb}                    & 85.52\%    & 85.57\% & \textbf{85.89}\% \\ \hline
\textbf{Telecom}                  & 87.81\%      & 89.11\% & \textbf{89.45}\% \\ \hline
\end{tabular}
\end{table}

\begin{figure}
    \centering
    \includegraphics[width=0.8\linewidth]{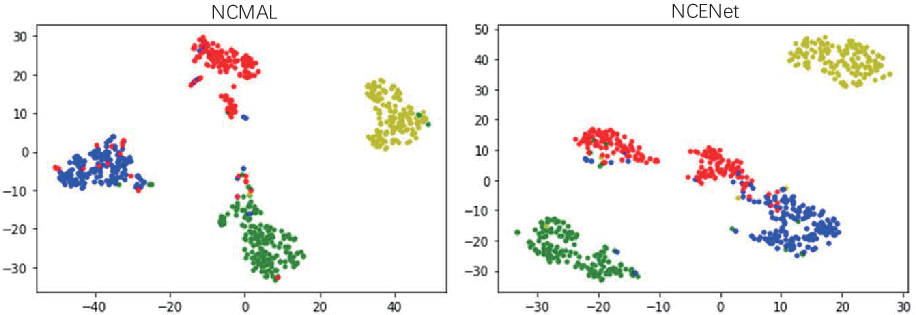}
    \caption{Demonstration of sample distribution on \textbf{AGNEWS} dataset using NCMAL and NCENet \eat{In NCENet, A portion of the red class is mixed with the blue class and the inter-class spacing is not obvious. Meanwhile, the intra-class compactness of the red class is also not ideal. In NCMAL, both inter-class discrepancy and intra-class compactness is satisfying}}
    \label{fig:comparison}
\end{figure}

\textbf{\textit{Performance on full training.}} \eat{To demonstrate the effectiveness of our approach, w}We first tested the performance of our model under full training with all samples labeled. This result is also equivalent to the test result under AL scenario at 100\% sample labeled. Table \ref{tab:ful-tr} illustrates the accuracy performance of our NCMAL and baselines fully trained on three datasets. Our NCMAL outperforms all baselines on all three datasets. This demonstrates the structural advantage of our classifier over both traditional softmax classifier and NCENet. Fig.\ref{fig:comparison} visualises the sample distribution of NCMAL and NCENet after full training on \textbf{AGNEWS} dataset. The classes formed by NCMAL are more compact compared to NCENet, which can be easily seen from the red class. At the same time, we can see from the distribution of blue class and red class that, the inter-class discrepancy are better assured by the NCMAL.

\begin{figure}[]
\centering
\subfigure[AGNEWS]{
\begin{minipage}[t]{0.33\linewidth}
\centering
\includegraphics[width=1.6in]{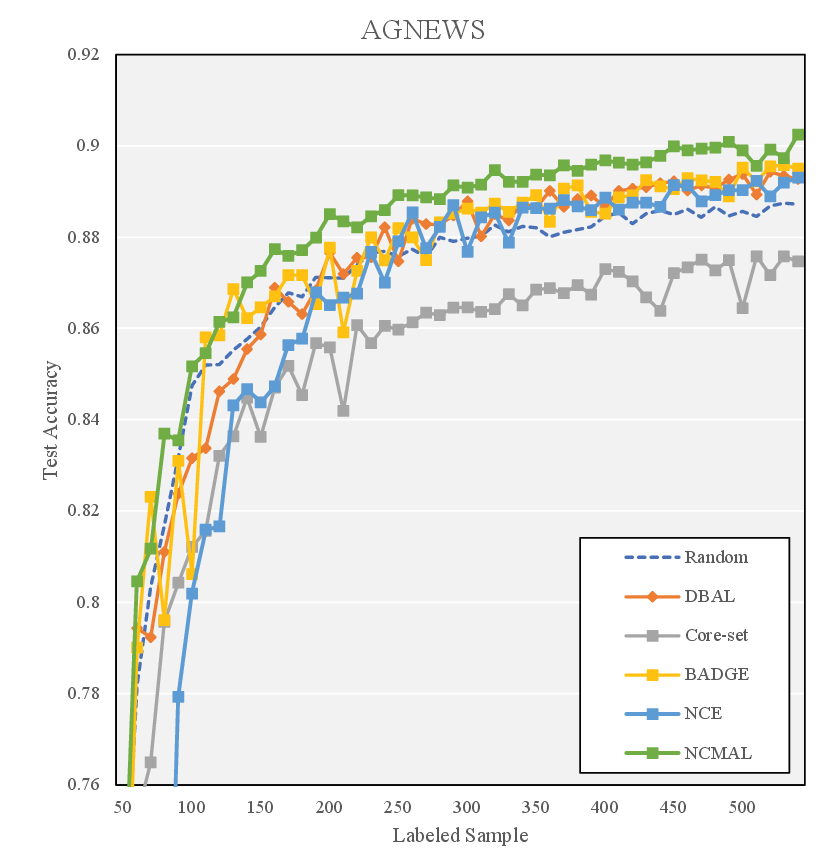}
\label{fig:hyp:hs}
\end{minipage}%
}%
\subfigure[IMDb]{
\begin{minipage}[t]{0.33\linewidth}
\centering
\includegraphics[width=1.6in]{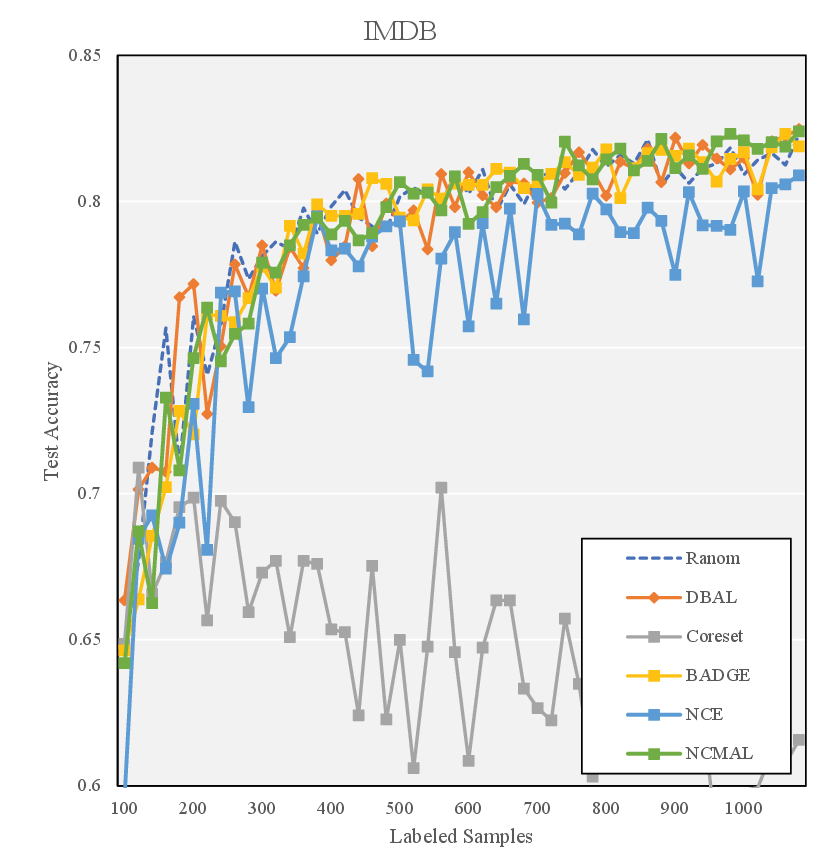}
\label{fig:hyp:gru}
\end{minipage}%
}%
\subfigure[Telecom]{
\begin{minipage}[t]{0.33\linewidth}
\centering
\includegraphics[width=1.6in]{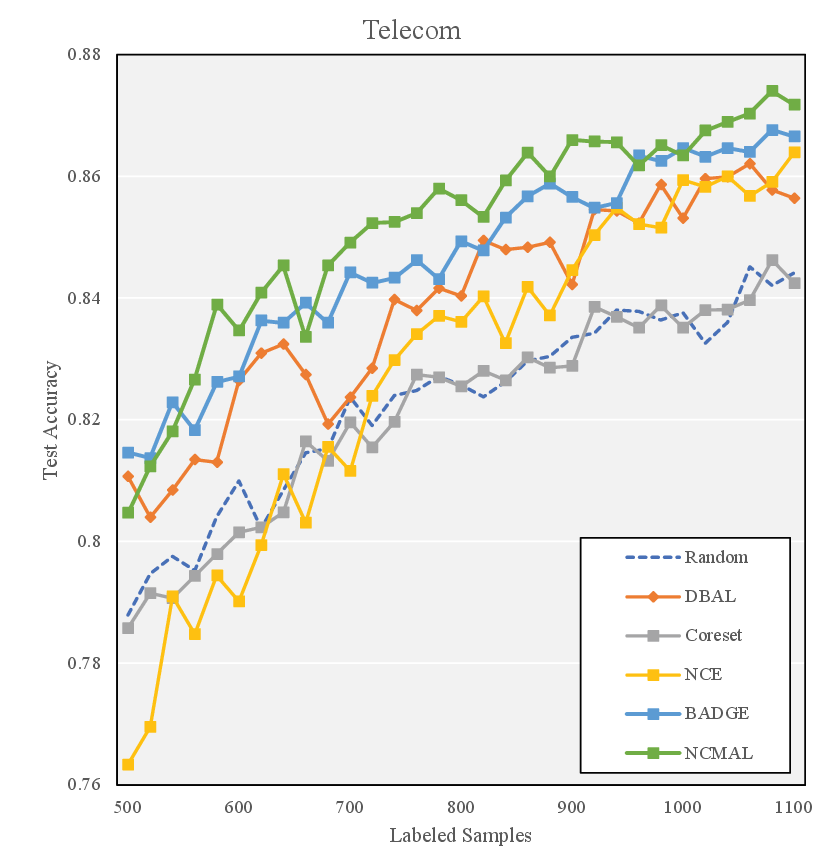}
\label{fig:hyp:gat}
\end{minipage}
}
\centering
\caption{Active learning performance and comparison with baseline methods}
\label{fig:hyp}
\end{figure}

\textbf{\textit{Performance on AL. }} Fig.\ref{fig:hyp} illustrates test accuracy curves of our NCMAL and baselines under AL scenario on three datasets. From Fig. \ref{fig:hyp}, we can draw three following critic conclusions.

    \emph{1)} In most cases except for Coreset, AL methods outperforms random selection, which shows the importance of AL methods. Meanwhile, our NCMAL outperforms all other baseline methods. Significant performance gaps can be observed on \textbf{Telecom} and \textbf{AGNEWS} dataset, in which NCMAL has a large performance gap with other methods from beginning to end. Under \textbf{IMDb} dataset, though advantages of all active learning methods over random selection are not very clear, our NCMAL still shows comparable performance over other baseline methods.
    
    \emph{2)} From Fig.\ref{fig:hyp} we can conclude that, the improvement are ascending as class number increases. Specifically, our method shows the greatest improvement over other methods on the \textbf{Telecom} dataset (25 classes) and marginal improvements on \textbf{IMDb} dataset (2 classes), which implies that, our methods is more suitable for classification with more classes. The reason may be that in classification with more classes, discrepancy between classes are even less guaranteed, and adding a margin term in such scenarios is more helpful in improving classification accuracy than in classification tasks where there are relatively few classes.
    
    \emph{3)} Compared with NCENet, our NCMAL shows constantly better performance especially in the front part of the learning curve. We believe that the addition of margin introduces a prior-knowledge to the model that classes are mutual exclusive, and thus the performance during early training period is improved.

\subsection{Differenct Sample Selection Strategies}

\begin{figure}[htbp]
\centering
\subfigure[AGNEWS]{
\begin{minipage}[t]{0.33\linewidth}
\centering
\includegraphics[width=1.6in]{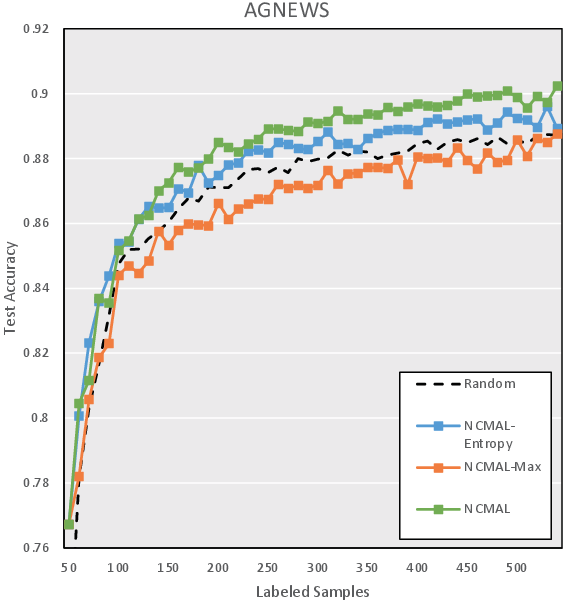}
\label{fig:hyp:hs}
\end{minipage}%
}%
\subfigure[IMDb]{
\begin{minipage}[t]{0.33\linewidth}
\centering
\includegraphics[width=1.6in]{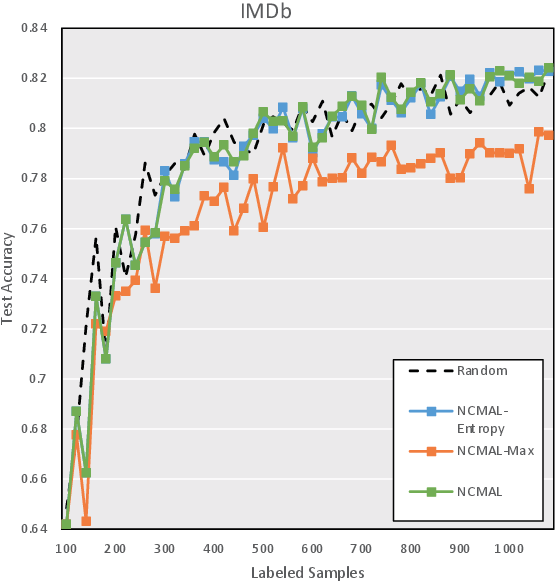}
\label{fig:hyp:gru}
\end{minipage}%
}%
\subfigure[Telecom]{
\begin{minipage}[t]{0.33\linewidth}
\centering
\includegraphics[width=1.6in]{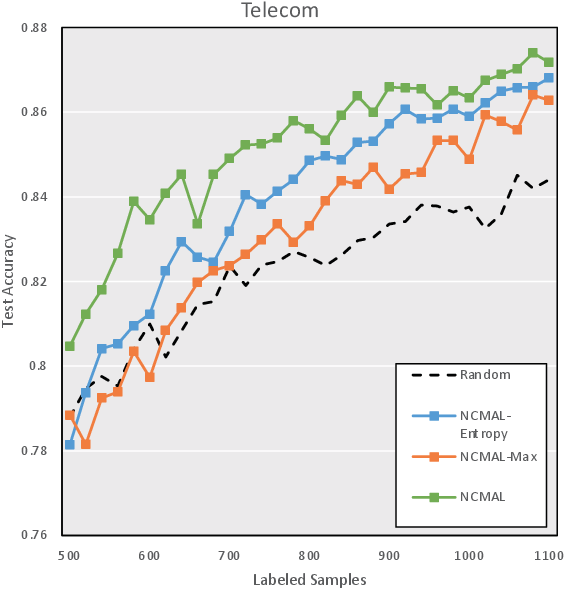}
\label{fig:hyp:gat}
\end{minipage}
}
\centering
\caption{Active learning performance and comparison with different variants of NCMAL}
\label{fig:ss}
\end{figure}

\begin{figure}
\centering
\subfigure[Margin Confidence]{
\begin{minipage}[t]{0.3\linewidth}
\centering
\includegraphics[width=1.5in]{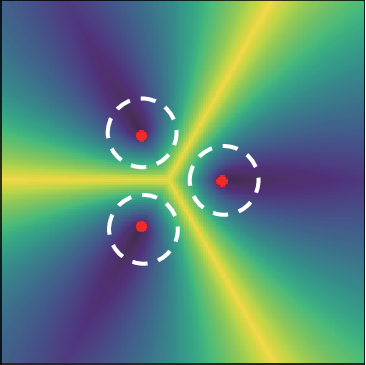}
\label{fig:samp:margin}
\end{minipage}%
}%
\subfigure[Entropy Confidence]{
\begin{minipage}[t]{0.3\linewidth}
\centering
\includegraphics[width=1.5in]{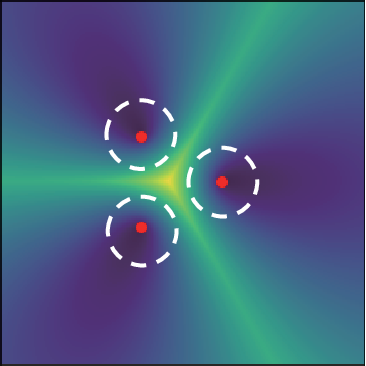}
\label{fig:samp:ent}
\end{minipage}
}
\subfigure[Max Confidence]{
\begin{minipage}[t]{0.3\linewidth}
\centering
\includegraphics[width=1.5in]{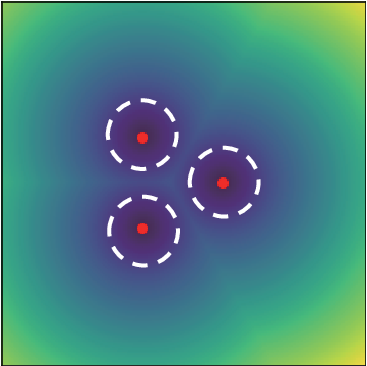}
\label{fig:samp:max}
\end{minipage}
}
\centering
\caption{Demonstration of the confidence score distribution of the two sample selection strategies. The red points represent the class centroids. The white dashed line represents the class classification boundary. Yellower areas have a higher confidence score, and bluer areas represent lower confidence score. The higher the confidence score, the higher the priority when labeling\eat{ In Margin Confidence, more samples from the margin area will be selected and labeled. In Entropy Confidence, more samples from center of class centroids will be selected and labeled. In Max Confidence, no distinct priority is given to samples in the margin area.}}
\label{fig:samp}
\end{figure}

With the same NCMAL network structure, the effect of sample selection strategies other than Margin Confidence is also studied. \eat{ExceptAs we can see from Fig.\ref{fig:samp}, the NCMAL-Margin method outperforms NCMAL-Entropy under nearly all circumstances. Except for NCMAL with Margin-Confidence (NCMAL for simplicity), two other As shown in Fig.\ref{fig:samp}, the difference between the two sampling strategies mainly lies in the different confidence score assigned to the margin area. NCMAL-Margin assigns a higher priority to the margin area and more samples in the margin are selected, and as a result better performance under AL scenario are gained. Since the more samples in the margin area selected, the better the AL performance, it means that the samples within the margin is more informative compared to other samples. This result supports our hypothesis presented in the previous section that samples in the margin area usually has more information and should have a higher priority when labeling. } Except for Margin Confidence (NCMAL for simplicity consideration), we bring out two variants of NCMAL with different sample selection strategies as comparison.
\vspace{-\topsep}
\begin{itemize}
\item \textbf{NCMAL-Entropy} 
To measure the confusion of the classifier, we propose a Entropy Confidence sample selection strategy.
\begin{equation}
    C^{Entropy}_x = \sum _c \left ( 1+o_{x,c} - \underset{c}{\mathrm{max}} o_{x,c} \right )
\end{equation}

\item \textbf{NCMAL-Max}
As the traditional uncertainty-based methods do, we pick the sample with the lowest maximum probability to be labeled.
\begin{equation}
    C^{Max}_x = - \underset{c}{\mathrm{max}} o_{x,c}
\end{equation}

 \end{itemize}
\vspace{-\topsep}
Those two sample selection strategies or their variants are often used in uncertainty-based methods. 

Fig.\ref{fig:ss} shows the test accuracy curve of NCMAL with three different sample selection strategies. It is easy to see that the original NCMAL with Margin Confidence has a significant performance advantage over the other two methods (NCMAL-Entropy \& NCMAL-Max) in most cases. Among them, NCMAL-Max performs the worst with weaker performance than random on both \textbf{AGNEWS} and \textbf{IMDb} data sets. Meanwhile, NCMAL-Entropy performs moderately, outperforming the random method on all three data sets and second only to original NCMAL.

We analyzed the reasons for this difference in performance. As shown in Fig.\ref{fig:samp}, the confidence score distribution of three sample selection strategies are demonstrated. The order of attention given to the margin area samples by different strategies is consistent with the order of their performance. The higher the confidence score given to samples from the margin area, the better the classification performance of the model obtains. The Margin Confidence in NCMAL gives the highest confidence score to samples from the margin area, thus more samples from the margin area will be selected and labeled. And as a result, NCMAL obtains the best performance. The Entropy Confidence gives moderate priority to samples in the margin area while Max Confidence gives nearly no priority to samples in the margin area, and as a result, NCMAL-Entropy gains moderate performance and NCMAL-Max obtains the worst performance. It is easy to see that the more samples from the margin area are selected, the better the model performs. This finding supports the hypothesis we presented in the previous section that, samples from the margin area tend to be more informative than samples from other areas and should have high priority when labeling. The combination of our NCMAL and Margin Confidence can best discover informative samples from the margin area, thus gains significant performance improvement.

\section{Conclusion}
In this paper, we propose a novel nearest neighbour classifier with margin for active learning (NCMAL). We add angular margin penalties so that the inter-class discrepancy can be assured thus the sigmoid classifier structure can be applied to mutual exclusive classification scenarios. This solves the problem of overlapping class boundaries that can occur in \cite{wana2021nearest} and achieves both better classification results and active learning results. We demonstrate the effectiveness of our method by comparing it with several baseline methods on different real datasets. We also proposed a special sample strategy in order to discover informative samples which lies in the margin area. The experimental results proves the correctness of our hypothesis and demonstrate the superiority of our method for text classification tasks. 

\section*{Acknowledgement}
This work was supported by National Natural Science Foundation of China (Grant No. 61702043, No.72274022).

%
%
%
%
\bibliographystyle{splncs04}

\end{document}